\documentclass[
 reprint,
superscriptaddress,
 amsmath,amssymb,
 aps,
 pra,
 floatfix,
]{revtex4-2}
\usepackage{graphicx}% Include figure files
\usepackage{dcolumn}% Align table columns on the decimal point
\usepackage{amsmath}
\usepackage{bm}% bold math
\usepackage[T1]{fontenc}
\usepackage{hyperref} % For hyperlinks

\DeclareMathOperator\erf{erf}

\newcommand{\ldash}[1]{\;\!---\;\!}

\begin{document}

\preprint{APS/123-QED}

\title{Cavity-induced quantum droplets}
\author{Leon Mixa}
\email{leon.mixa@uni-hamburg.de}
\affiliation{I. Institut für Theoretische Physik, Universität Hamburg, Notkestraße 9, 22607 Hamburg, Germany}
\affiliation{The Hamburg Center for Ultrafast Imaging, Luruper Chaussee 149, 22761 Hamburg, Germany}
\author{Milan Radonji\'c}
\affiliation{I. Institut für Theoretische Physik, Universität Hamburg, Notkestraße 9, 22607 Hamburg, Germany}
\affiliation{Institute of Physics Belgrade, University of Belgrade, Pregrevica 118,
11080 Belgrade, Serbia}
\author{Axel Pelster}
\affiliation{Physics Department and Research Center OPTIMAS, Rheinland-Pfälzische Technische Universität Kaiserslautern-Landau, Erwin-Schrödinger Straße 46,
67663 Kaiserslautern, Germany}
\author{Michael Thorwart}
\email{michael.thorwart@uni-hamburg.de}
\affiliation{I. Institut für Theoretische Physik, Universität Hamburg, Notkestraße 9, 22607 Hamburg, Germany}
\affiliation{The Hamburg Center for Ultrafast Imaging, Luruper Chaussee 149, 22761 Hamburg, Germany}

\date{\today}

\begin{abstract}
Quantum droplets are formed in quantum many-body systems when the competition of quantum corrections with the mean-field interaction yields a stable self-bound quantum liquid. We predict the emergence of a quantum droplet when a Bose-Einstein condensate is placed in a transverse-pumped optical cavity. The strong coupling between the atoms and a cavity mode induces long-range interactions in the atoms, and a roton mode for negative cavity detuning emerges. Using Bogoliubov theory, we show that the roton mode competes with the repulsive atomic $s$-wave scattering. 
Due to the favorable scaling of the quantum fluctuations with respect to the volume, a self-bound stable quantum liquid emerges.
\end{abstract}

\maketitle

\section{Introduction}
Superfluid helium-4 is known for intricate macroscopic quantum states, the most famous of which are probably the superfluid states \cite{kapitza1938viscosity} beside the roton mode in its dispersion relation \cite{landau1941theory}.
It also realizes a state of zero pressure at a specific density as a free droplet state \cite{volovik2003universe}.
Since liquid helium is highly correlated, the long-range correlations and the underlying quantum many-body interactions are not easy to reveal.
Progress has been made by constructing effective energy functionals \cite{stringari1987systematics,dupont1990inhomogeneous,casas1995density,dalfovo1995structural} and benchmarking them against experimental data rather than from first principles. 
The fact that a quantum liquid of particle number $N$ and volume $V$ is self-confined in free space can ultimately be reduced to three conditions that its effective energy $E_0(N,V)$ has to fulfill \cite{volovik2003universe}.
First, (C1) states that a local extremum defined by $(\partial E_0/\partial V)_N = 0$ determines the equilibrium system volume $V_0$.
This becomes the zero-pressure condition in the thermodynamic limit.
For stability, the local extremum must be a minimum, so condition (C2) reads $(\partial^2 E_0/\partial V^2)_N |_{V=V_0} > 0$ and implies a positive bulk compressibility in the thermodynamic limit.
Finally, the droplet should not self-evaporate, so condition (C3) amounts to $(\partial E_0/\partial N)_{V}|_{V=V_0} < 0$, implying a negative effective chemical potential.
There has to be a competition of both attractive and repulsive contributions, depending differently on the system size $V$, to satisfy conditions (C1)$-$(C3). An effective toy model energy is given by $E_0(N,V) = \alpha(N) V^{-1} + \beta(N) V^{-1-\gamma}$, where the first term can be the attractive mean-field energy, while the second is the repulsive quantum fluctuation correction.
It turns out that three choices of the parameters $\alpha$, $\beta$, and $\gamma$ can satisfy the droplet conditions (C1) and (C2). 
They are (D1) $\alpha < 0$, $\beta > 0$, $\gamma>0$, (D2) $\alpha>0$, $\beta<0$, $0>\gamma>-1$, and (D3) $\alpha >0$, $\beta>0$, $\gamma<-1$.
Whether the third condition (C3) is also satisfied turns out to depend on the specific $N$-dependence of $\alpha$ and $\beta$.
An additional volume-independent term $E^{(\infty)}(N)$ in the effective energy $E_0$ can help satisfy (C3) and arises naturally from the infinite-range interaction considered in this article.
\\
\indent
Recently, quantum droplets have been identified as a unique quantum state of matter.
In contrast to helium droplets, they are formed in weakly interacting dilute atom gases.
Their discovery in Bose-Bose mixtures was initiated by the seminal prediction of Petrov \cite{petrov2015quantum} that the competition between intra- and inter-species contact interactions can cause the mean-field energy functional to become unstable.
Simultaneously, the zero-point motion of the Bogoliubov excitations of the mixture, as captured by the Lee-Huang-Yang (LHY) corrections \cite{lee1957eigenvalues},  stabilizes the system.
For a given particle density $n=N/V$, the competition between the unstable mean-field energy $\propto n^2$ and the stabilizing LHY term $\propto n^{5/2}$ establishes an equilibrium for which the system pressure vanishes for finite $n$ in the thermodynamic limit.
Such a droplet realization is of the type (D1).
These self-bound quantum droplets of finite size are described by an extended Gross-Pitaevskii equation, where the LHY corrections are included via a local density approximation \cite{petrov2015quantum}.
Experimental realizations were not only reported in Bose-Bose mixtures \cite{cabrera2018quantum,semeghini2018self,skov2021observation} but likewise in dilute Bose gases with dipolar interaction \cite{ferrier2016observation,pfau-nature,chomaz2022dipolar}. 
The extended Gross-Pitaevskii equation including the dipolar LHY correction \cite{schutzhold2006mean,lima2011quantum,lima2012beyond} also serves as the theoretical framework for their description \cite{wachtler2016quantum,bisset2016ground}.
Importantly, dipolar Bose-Einstein condensates (BECs) host the roton mode, which was predicted in 2003 \cite{roton-prediction1,roton-prediction2} and later verified experimentally \cite{chomaz2018observation,petter2019probing}.
Besides three-dimensional quantum droplets, two- and one-dimensional ones are also studied \cite{Petrov2016,Ilg,jia2022expansion}.
One-dimensional mixture droplets represent an effective model corresponding to (D2) \cite{Petrov2016}.
\\
\indent
When a three-dimensional BEC is placed in an optical cavity, the Dicke model of cavity QED is realized \cite{dicke1954coherence,nagy2010dicke,baumann2010dicke}.
A transverse pump beam off-resonantly drives internal atomic transitions.
The quantum states of the cavity and the atoms are strongly coupled by the scattering of pump photons into the cavity and the repeated scattering of cavity photons by the atoms \cite{mivehvar2021cavity}. 
The cavity thus mediates a long-range effective atom-atom interaction.
For sufficiently strong coupling, self-organization and the Dicke quantum phase transition are triggered \cite{nagy2010dicke,baumann2010dicke,dicke1954coherence,emary2003chaos} together with a softening of a roton-like mode of the atom gas \cite{mottl2012roton}.
Cavity-induced droplet phases have been studied for an ultracold gas held in an external optical lattice, realizing the extended Bose-Hubbard model.
Droplets emerge by sign-changing potentials competing with entropy effects \cite{karpov2019crystalline}, or by an interplay of on-site and cavity-induced long-range interactions \cite{karpov2022light}.
\\
\indent
In this article, we show that the quantum fluctuations of the roton mode give rise to the formation of quantum droplets.
We show that it is crucial to consider the finite range of the cavity-mediated interaction, which is usually assumed to be infinite.
The mechanism differs in several aspects from that observed in either a mixture or a dipolar Bose gas. 
There, the contact atom-atom interactions are tuned via Feshbach resonances so that the system becomes unstable at the mean-field level and the atomic quantum fluctuations stabilize the macroscopic state, corresponding to either (D1) or (D2).
In contrast, cavity quantum droplets formed in a mean-field stable cavity BEC rather are of type (D3). They are formed for experimentally realistic densities due to the impact of quantum fluctuations of the cavity mode. The concept developed in this work for the case of the cavity-induced long-range interaction can be applied to a broader class of systems with finite interaction range \cite{companionPaper}.

\section{Cavity BEC model}
The experimental setup is shown in the inset of Fig.\ \ref{energyPerParticle}.
A BEC of density $n$ is placed in the center of a cavity.
Each atom of mass $M$ is considered as a two-level system $|g\rangle \leftrightarrow |e\rangle$ with frequency separation $\omega_{\rm A}$.
A pump beam of frequency $\omega_{\rm P}$ along the $y$-axis drives the transition with a large detuning {$\Delta_{\rm A} = \omega_{\rm P} - \omega_{\rm A}$}.
The beam has the spatially dependent Rabi frequency {$h(\bm{r}) = h_0 \cos(ky)$}, where $k$ is the wavenumber of the pump photons. 
Due to Rayleigh scattering into the cavity, which is enhanced by the Purcell effect, the atoms are coupled to a quantized cavity mode of frequency $\omega_{\rm C}$ along the $x$-direction.
The cavity is red detuned by {$\Delta_{\rm C} = \omega_{\rm P} - \omega_{\rm C} < 0$} and damped at the rate $\kappa$.
The maximal Rabi frequency ${\cal G}_0$ determines the effective coupling strength {$U_0 = {\cal G}_0^2/\Delta_{\rm A}$} between an atom in the ground state $|g\rangle$ and the cavity.
We consider the TEM$_{00}$ mode {${\cal G}(\bm{r}) = {\cal G}_0 \cos (kx) \exp[-(y^2+z^2)/\xi^2]$} with the Gaussian envelope of waist $\xi$ transverse to the cavity axis.
This results in a spatially dependent atom-cavity interaction \cite{maschler2008ultracold}.
Physically, the pump mode function $h(\bm{r})$ also has a transverse Gaussian profile, 
which is neglected, but does not alter qualitatively the results. 
Experimentally, this simplification corresponds to a broad pump beam profile requiring external confinement along the cavity axis.
The $s$-wave scattering length $a_s$ of the atoms in the ground state determines their contact interaction strength {$g=4\pi a_s/M$} with {$\hbar = 1$}.
We assume the hierarchy of magnitudes $|\Delta_{\rm A}| \gg |\Delta_{\rm C}| \gg k^2/(2M) \gg gn, \, |U_0|$, consistent with contemporary experimental setups.
This allows to eliminate both the excited atomic state $|e\rangle$ and the dissipative fast cavity mode \cite{maschler2008ultracold,mivehvar2021cavity}. After resolving the ambiguities in the operator ordering \cite{jager2022lindblad}, an effective atom-only Hamiltonian in terms of the atomic ground state field operator $\hat{\psi}(\bm{r})$  follows as
\begin{align}
&\hspace*{-2mm}\hat{H}_{\text{eff}} = \int_V d^3\bm{r}\;\! \hat{\psi}^{\dag}(\bm{r}) \left[ -\frac{\bm{\nabla}^2}{2M} + \frac{g}{2}\;\! \hat{\psi}^{\dag}(\bm{r}) \hat{\psi}(\bm{r}) \right]\hat{\psi}(\bm{r}) \nonumber \\
&\hspace*{-1mm}{}+{}\frac{1}{2} \int_V d^3\bm{r}\int_V d^3 \bm{r}' \hat{\psi}^\dag(\bm{r}) \hat{\psi}(\bm{r}) V_{\rm C}(\bm{r},\bm{r}') \hat{\psi}^{\dag}(\bm{r}') \hat{\psi}(\bm{r}') \, .
\label{effectiveInteractingBEC}
\end{align}
The cavity-induced atom-atom interaction is given by
\begin{align}
V_{\rm C}(\bm{r},\bm{r}') ={}&\mathcal{I} \cos{(kx)} \cos{(ky)}\, e^{- (y^2+z^2)/\xi^2} \nonumber \\
&\times \cos{(kx')} \cos{(ky')} \, e^{-(y'^2+z'^2)/\xi^2} \, ,
\label{cavityInteractionPotential}
\end{align}
where cavity and pump parameters lead to the effective interaction strength $\mathcal{I} = 2 {\cal G}_0^2 h_0^2 \Delta_{\rm C}/[\Delta_{\rm A}^2(\Delta_{\rm C}^2+\kappa^2)] < 0$. 

\subsection{Homogeneous mean-field approach}
We assume that the atoms occupy a volume $V$ that is a cube of size $L<\xi$. It is straightforward to generalize the calculation to other shapes.
Below the superradiant Dicke phase transition, the cavity remains empty and the condensate is homogeneous at the mean-field level \cite{nagy2011critical}, so that we can replace $\hat{\psi}(\bm{r})\to\langle \hat{\psi}(\bm{r}) \rangle = \sqrt{n}$. 
The interaction potential of Eq.\ (\ref{cavityInteractionPotential}) leads to integrals of the type
\begin{align}
\int_{-L/2}^{L/2} du\;\! e^{iku} e^{-u^2/\xi^2} \approx \delta_{k0} \sqrt{\pi}\xi\erf\left( \frac{L}{2\xi} \right) 
\label{KroneckerDelta}
\end{align}
in the evaluation of both the mean-field and the quantum fluctuations below,
where the approximation is derived in Appendix \ref{App:A}.
Note that the above result involves a finite interaction range $\xi$ and a finite system extent $L$.
Thus the {\it finite-size} property of the system is a direct consequence of the finite interaction range $\xi > L$, while the number of atoms $N$ must be physically finite.
The condensate mean-field energy reads
\begin{align}
E_{\text{mf}} = V \frac{g}{2} n^2 \, .
\label{meanFieldPotential} 
\end{align}
The homogeneous mean-field condensate and its stability are unaffected by the cavity below the Dicke transition.

\subsection{Quantum fluctuations}
Next we develop a Bogoliubov theory of the fluctuations $\hat{\phi}(\bm{r}) = \hat{\psi}(\bm{r}) - \sqrt{n}$ around the homogeneous mean-field using the plane-wave expansion and 
Eq.\ (\ref{KroneckerDelta}) for the cavity-induced interaction. 
We defer the technical discussion to Appendix \ref{App:B}.
Importantly, within the approximation leading to Eq.\ (\ref{KroneckerDelta}), the cavity couples only to the four Bogoliubov modes \mbox{$\mathcal{K}_{\rm C} = \lbrace ({\!}\pm{\;\!\!}k\;\:\pm{\;\!\!}k\;\: 0 )^T \rbrace$} determined by the wavenumber $k$ of the light field.
The remaining atomic modes not coupled to the cavity have the standard Bogoliubov dispersion $\omega_{\bm{p}} = \sqrt{ \bm{p}^2/(2M) [\bm{p}^2/(2M) + 2gn ]}$ in accordance with the Goldstone theorem.
Their zero-point fluctuations in the continuum limit yield the LHY energy correction of the Bose gas \cite{lee1957eigenvalues}. 
For a dilute weakly interacting Bose gas with $a_s^3 n\ll 1$, the LHY correction is negligible compared to the mean-field energy, Eq.\ (\ref{meanFieldPotential}).
\\
\indent
The four modes of $\mathcal{K}_{\rm C}$ form a set of mutually interacting quantum harmonic oscillators with eigenfrequencies $\{ \omega_{\bm{k}},\omega_{\bm{k}},\omega_{\bm{k}},\Omega \}$ as described by Eqs.\ (\ref{qfHamiltonian} - \ref{couplingMatrix}), where
\begin{align}
\Omega = \sqrt{\omega_{\bm{k}}^2 + 16 \frac{k^2}{M} V_{\rm C}(\bm{k},\bm{k}) n} 
\label{rotonMode}
\end{align}
denotes the frequency of the single resulting roton mode.
It involves the Fourier transform of the cavity-mediated interaction potential, which according to Eq.\ (\ref{KroneckerDelta}) reads
\begin{align}
\hspace*{-2mm}
V_{\rm C}(\bm{p},\bm{p}') = \frac{\mathcal{I}V}{32} \left[ \frac{\sqrt{\pi}\xi}{L} \erf\left( \frac{L}{2\xi} \right) \right]^4 \sum_{\bm{k},\bm{k}'\in\mathcal{K}_C} \delta_{\bm{p}\bm{k}} \delta_{\bm{p}'\bm{k}'} \, .
\label{cavityFourierTransform}
\end{align}
The roton softness is determined by the effective cavity-induced interaction strength $\mathcal{I}$.
It can be tuned by changing the pump strength $h_0$ or the cavity detuning $\Delta_{\rm C}$, but can also be fine-tuned by adopting any other cavity feature.
The roton mode is responsible for the leading-order energy contribution derived as Eq.\ (\ref{quantumCorrection}) of Appendix \ref{App:B}
\begin{eqnarray}
E_{\text{ac}} = \frac{1}{2} \left( \Omega - \omega_{\bm{k}} \right) \, .
\label{cavityCorrection}
\end{eqnarray}
It corresponds to the difference between the respective energies with and without coupling to the cavity and vanishes in the limit $\mathcal{I}\to 0$.
Combining Eq.\ (\ref{meanFieldPotential}) with Eq.\ (\ref{cavityCorrection}) yields the effective beyond-mean-field energy $E_0 = E_{\rm mf} + E_{\text{ac}}$ of the condensate ground state.
The chemical potential then follows
as $\mu_0 = g n + (\partial E_{\text{ac}}/\partial N)_V$.

\section{Effective energy}
The formation of quantum droplets can be understood by analyzing $E_0$.
\begin{figure}
  \centering
  \includegraphics[scale=.6]{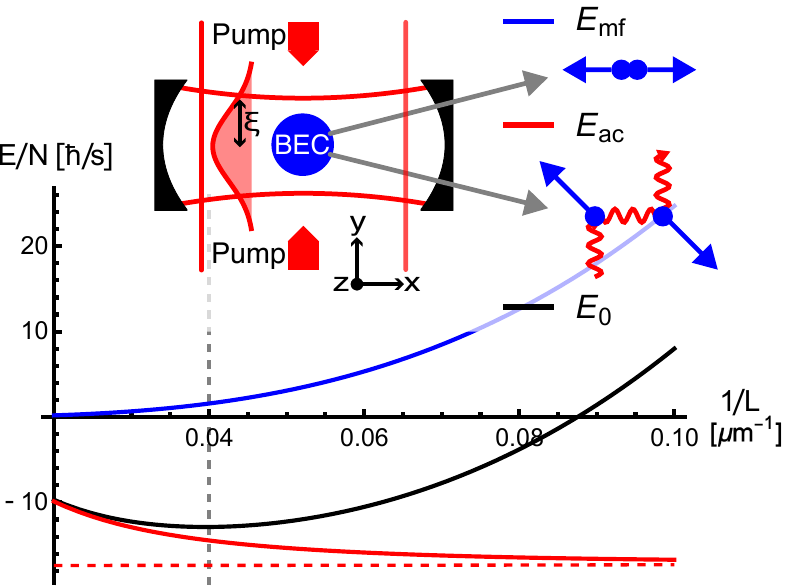}
  \caption{
  Effective ground-state energy per atom $E_0/N$ with repulsive contact interaction mean-field ($E_{\text{mf}}$) and cavity-mediated quantum fluctuation contribution ($E_{\text{ac}}$) plotted against the inverse cloud size $1/L$ for a fixed atom number $N=10^3$.
 Cartoons depict the contact and the cavity-mediated interaction.
 The vertical dashed line marks energy minimum corresponding to the equilibrium droplet size $L_0$. The red dashed line indicates $E_{\text{ac}}^{(\infty)}$ for an infinite-range cavity $L/\xi \to 0$. Other parameters are $\mathcal{I} = 0.95 \,\mathcal{I}_{\rm cr}$, $\xi=50\, \mu$m, $a_s = 100\, a_0$, and $M=87\,u$. Inset: Three-dimensional BEC (in blue) placed at the cavity center aligned along the $x$-axis with a Gaussian mode profile of waist $\xi$. The broad pump laser propagates along the $y$-direction. \label{energyPerParticle}}\vspace{-2mm}
\end{figure}
For fixed $N$, the conditions (C1) and (C2) impose that the system exhibits an energy minimum with respect to the volume $V$ \cite{gajda2021stability}.
For repulsive contact interaction, the BEC mean-field energy Eq.\ (\ref{meanFieldPotential}) is positive and convex with respect to $V$.
Conversely, due to $\Delta_{\rm C} < 0$ implying $\mathcal{I}<0$,
the contribution of the cavity quantum fluctuations in Eq.\ (\ref{cavityCorrection}) turns out to be negative, which is typical for a roton mode.
Thus, the combined effective energy $E_0$ can have a minimum as depicted in Fig.\ \ref{energyPerParticle}.
To maintain the homogeneous phase, we need to stay below the Dicke phase transition.
Since the latter occurs when the roton mode becomes soft, from the condition $\Omega=0$ we obtain the critical value $\mathcal{I}_{\rm cr} = -2[(k^2/M)+2gn]/[N(\sqrt{\pi}\erf[L/2\xi]\;\!\xi/L)^4]$.

The emergence of $E_{\text{ac}}$ from a single roton mode has an important physical implication.
The thermodynamic limit is $N, L, \xi \to \infty$, while $N/V =$ const and $L/\xi =$ const.  
Furthermore, the coupling of a single atom to the cavity vanishes, i.e., ${\cal G}_0 \to 0$, so that $\mathcal{I} V$ is constant. Then, the roton's energy contribution becomes intensive, i.e., induces a {\it finite-size effect}.
Consequently, in the limit of large $N$, the largest possible value of $E_{\text{ac}}$, at $\Omega = 0$, will eventually be dwarfed by any extensive energy term.
Hence, self-bound droplets are only possible for systems of finite size.
\\
\indent
To gain further understanding, we expand $V_{\rm C}(\bm{p},\bm{p}')$ in Eq.\ (\ref{cavityFourierTransform}) around $L/\xi = 0$ to the second order, which is accurate for a realistic experiment with global interaction range $L/\xi<1$.
The cavity quantum fluctuation correction is then expanded as $E_{\text{ac}} \approx E_{\text{ac}}^{(\infty)} + D L^2/2$.
The first term $E_{\text{ac}}^{(\infty)}$ is the energy correction in the infinite interaction range limit $L/\xi \to 0$, indicated by the dashed horizontal line in Fig.\ \ref{energyPerParticle}.
It does not depend on $L$ and is equivalent to taking the zeroth-order expansion in $u/\xi$ of the Gaussian in Eq.\ (\ref{KroneckerDelta}), i.e., a constant transverse profile of the mode function ${\cal G}(\bm{r})$.
By this, we recover the known results for the infinite-range cavity-induced interaction \cite{nagy2011critical}.
The second term, resulting from the specific shape of the envelope, contains the factor $D = -\mathcal{I}N/\big[\:\!12\xi^2 \sqrt{1+(4gn+\mathcal{I}N)M/(2k^2)}\;\!\big]$.
We ignore the contact interaction $gn\ll k^2/M$ in $E_{\text{ac}}$, which is valid $\mathcal{I}$ is not too close to $\mathcal{I}_{\rm cr}$. The energy correction then has the form of a harmonic-like self-trapping potential $\propto L^2$. The effective energy becomes
\begin{align}
\hspace*{-3mm}
E_0(N,V) = E_{\text{ac}}^{(\infty)} + \frac{g N^2}{2 V} - \frac{\mathcal{I} N V^{2/3}}{24\xi^2\sqrt{1+\mathcal{I}NM/(2k^2)}} \hspace*{-2mm}
\label{effectiveModel}
\end{align}
and gives rise to the equilibrium droplet volume $V_0^{5/3} = -18 g N \xi^2 \sqrt{1 + \mathcal{I}NM/2k^2}/\mathcal{I}$. 
This corresponds to the minimal droplet model of type (D3) by identifying $\alpha = gN^2/2>0$, $\beta = D/2 >0$, and $\gamma = -5/3$.
\begin{figure}
  \centering
  \includegraphics[scale=.42]{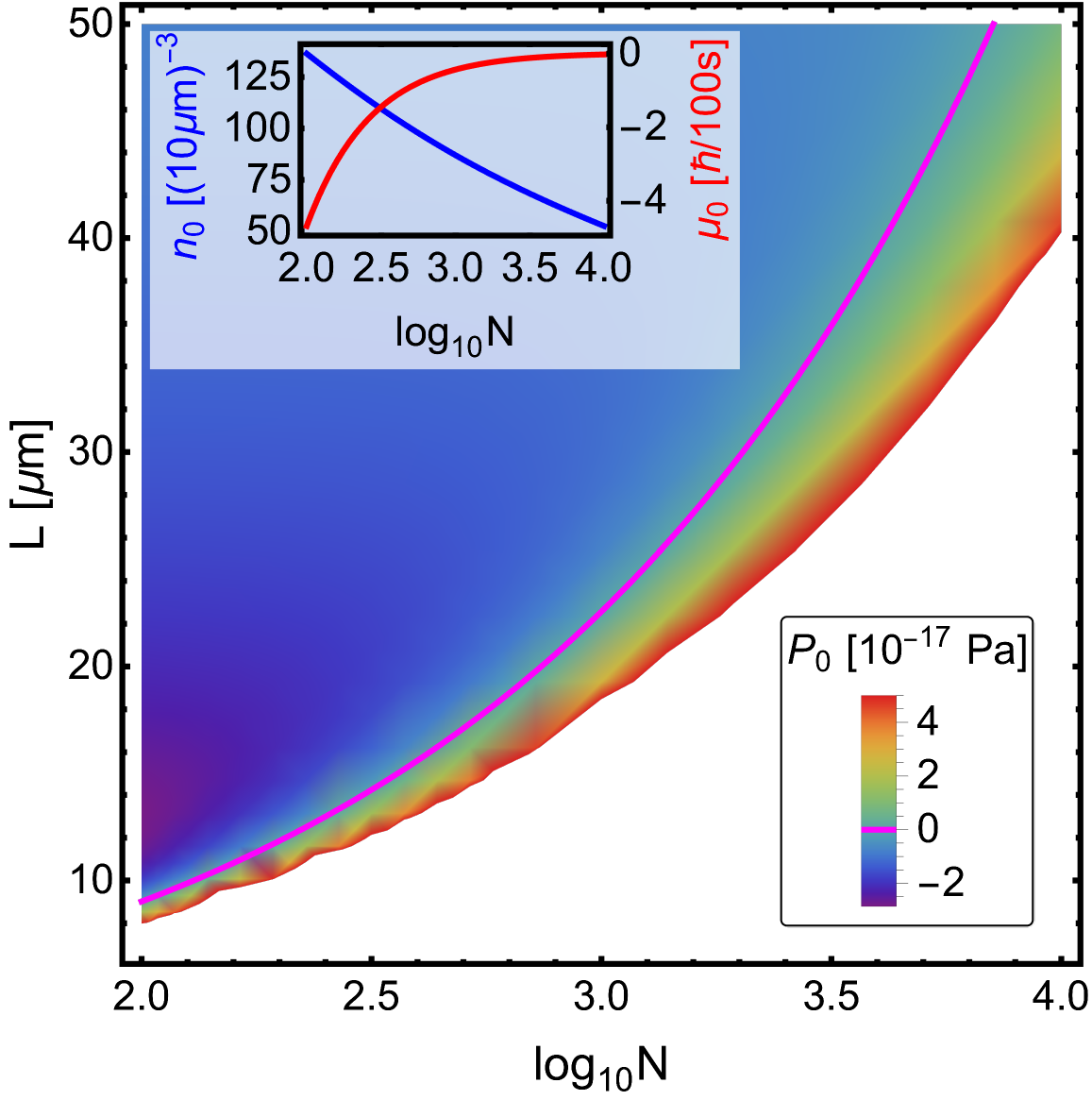}
  \caption{System pressure $P_0$ dependence on the atom number $N$ and the system size $L$ for $\mathcal{I}=0.95 \,\mathcal{I}_{\rm cr}$. The magenta curve is the zero-pressure droplet line. Inset: Droplet density (blue) and chemical potential (red) realized along this line. The absolute value of $\mathcal{I}_{\rm cr}\sim N^{-1}$ decreases from left to right. Remaining parameters as in Fig.\ \ref{energyPerParticle}.  
  \label{pressureDiagram}}
\end{figure}

\subsection{Thermodynamic considerations}
As stated in (C1) and (C2), energy minimization with respect to the system size translates into a zero pressure condition \mbox{$0 = -(\partial E_0/\partial V)_N = P_0$} with positive bulk compressibility \mbox{$0 < K(P_0 = 0) = -V (\partial P_0/\partial V)_N|_{V=V_0}$}, corresponding to a thermodynamically stable state \cite{volovik2003universe}.
The mean-field contact interaction yields a positive pressure $P_{\text{mf}} = gn^2/2$, while the roton mode corresponds to a negative pressure $P_{\text{ac}} = -D/(3L)$.
As the cavity-induced interaction becomes stronger with increasing $\mathcal{I}$, the negative pressure $P_{\text{ac}}$ becomes comparable to the positive mean-field pressure $P_{\text{mf}}$.
Their interplay results in a stable droplet with zero total pressure and positive bulk compressibility, as displayed in Fig.\ \ref{energyPerParticle}.
The compressibility also obeys $K(P_0 = 0)/n < gn$.
The corresponding sound velocity is modified by the roton mode contribution and turns out to be reduced with respect to the Bose gas counterpart.
\\
\indent
The system pressure is depicted in Fig.\ \ref{pressureDiagram} for varying particle number $N$ and system size $L$.
The magenta line marks $P_0 = 0$, where the droplet size $L_0 = V_0^{1/3}$ is determined by $N$ and other system parameters.
We find a negative pressure above this line and a positive pressure below it, indicating the positive bulk compressibility of the droplet corresponding to each number of atoms $N$.
As shown in Fig. \ref{pressureDiagram}, the droplet size depends nonlinearly on the number of atoms, in accordance with the finite size nature of the droplet.
The pressure gradient, related to the droplet compressibility, decreases as $N$ increases.
The inset shows the equilibrium density $n_{0}=N/V_0$ realized along the zero-pressure line.
As the number of particles increases, the droplet density decreases monotonically.
The decrease is mainly due to the fact that the cavity interaction $\mathcal{I} = 0.95 \,\mathcal{I}_{\rm cr}$ is kept fixed with $\mathcal{I}_{\rm cr}\propto N^{-1}$.
In principle, denser droplets can be realized by getting closer to the Dicke phase transition \cite{companionPaper}.
This is due to the divergence of the cavity pressure $P_{\text{ac}}$ when the roton mode vanishes as $\mathcal{I}\to\mathcal{I}_{\rm cr}$.
We note that this happens alongside the diverging condensate depletion into the roton mode.
In line with contemporary experiments, we have chosen to stay $5\%$ below the critical point. With this, we extract the leading order relation $V_0 \sim N^{3/5}$ from our analytical expression for $V_0$.
\begin{figure}
    \centering
    \includegraphics[scale=0.42]{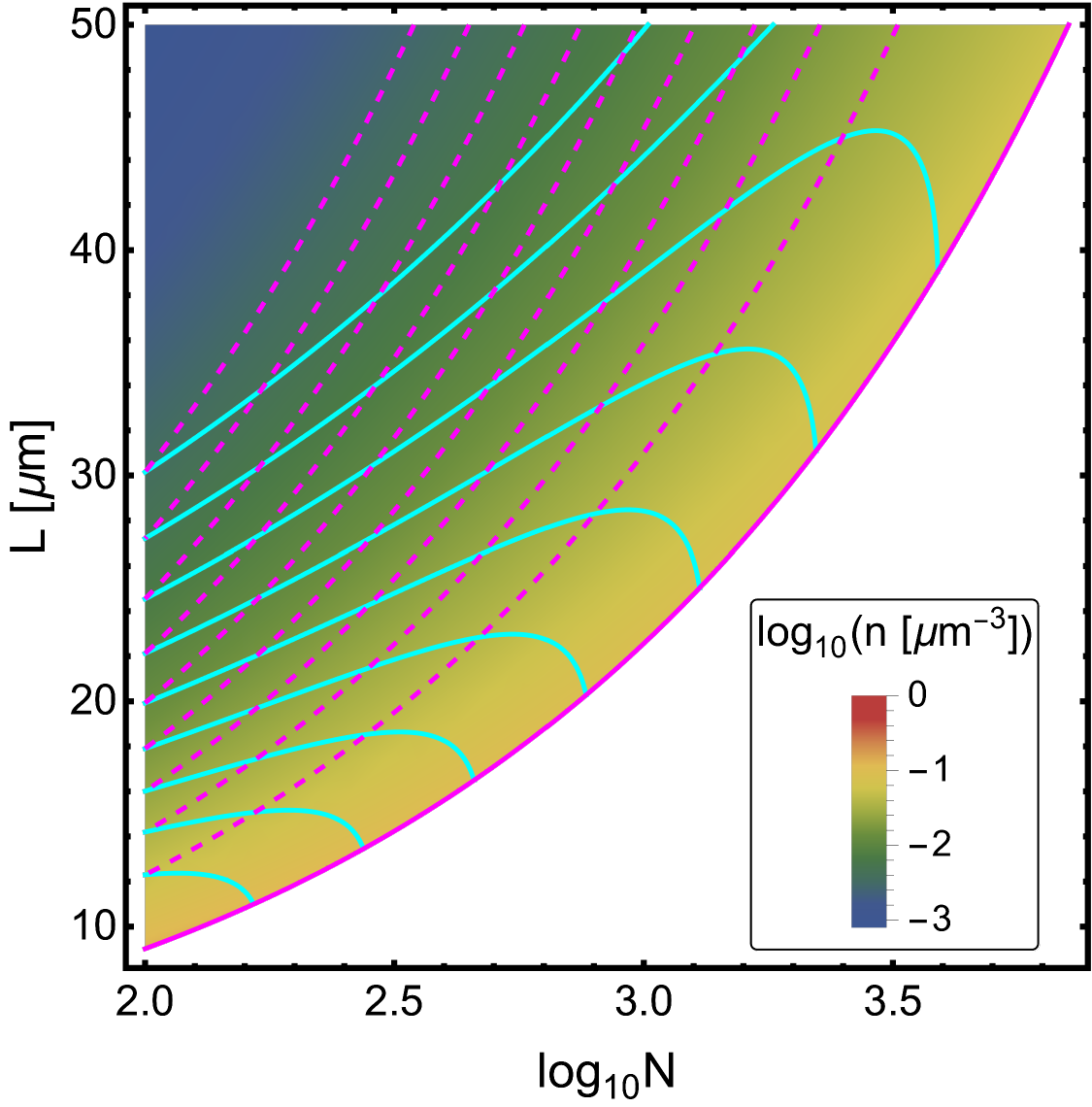}
    \caption{
    Droplet density on a logarithmic scale as a function of the atom number and the system size. Dashed and full lines correspond to zero-pressure contours as described in the text.
    Parameters are the same as in Figs.\ \ref{energyPerParticle} and \ref{pressureDiagram}.
    }
    \label{DensityInteraction}
\end{figure}

\subsection{Chemical potential}
Finally, we check whether the chemical potential remains negative \cite{volovik2003universe} such that the droplet does not evaporate spontaneously. 
The total chemical potential $\mu_0 = g n + (\partial E_{\text{ac}}/\partial N)_V$ is composed of the positive mean-field and the negative roton contribution.
At zero pressure, the cavity-induced term prevails over $gn$, resulting in a negative overall $\mu_0(P_0=0) < 0$, thus avoiding self-evaporation \cite{companionPaper}.
As displayed in Fig.\ \ref{pressureDiagram} (inset), the absolute value of the total chemical potential diminishes for growing $N$ in the same way as the density. Yet, it always remains negative.

\subsection{Droplet density}
Figure \ref{DensityInteraction} displays the droplet density as a function of the atom number and the system size.
The magenta lines correspond to the zero-pressure contours of long-range interaction strength values, $\mathcal{I}$, ranging from $0.1\,\mathcal{I}_{\rm cr}$ to $0.95\,\mathcal{I}_{\rm cr}$, from the top left to the lower right corner as a geometric sequence.
The magenta border line is the same as the magenta line in Fig.\ \ref{pressureDiagram}.
Recall that $\mathcal{I}_{\rm cr} \sim N^{-1}$ depends on the number of atoms.
The solid cyan lines correspond to zero-pressure contours with {\it constant} values of $|\mathcal{I}| \approx 9\, \text{ Hz\ldash{}}\, 515$ Hz, which are chosen to match the dashed curves for $N = 100$.
Quantitative examination of the results presented in Fig. \ref{DensityInteraction} reveals that compared to typical BEC atom numbers and other droplet realizations \cite{cabrera2018quantum,semeghini2018self,skov2021observation,ferrier2016observation,pfau-nature}, the droplet densities presented here are orders of magnitude more dilute.
This has its root in the BEC mean-field contact interaction that the cavity fluctuations have to overcome.
In other droplet realizations, the mean field is almost completely suppressed from the outset.
In addition, the homogeneous nature of the droplets is constrained by the Dicke phase transition. Droplets of any desired density can be obtained by approaching the Dicke critical point, as shown in Fig. \ref{DensityInteraction}.
However, due to technical limitations in experimental control, it is more realistic to optimize other system parameters, such as, e.g., the equilibrium density that scales approximately as $n_{0}\sim g^{-3/5}$ with respect to the contact interaction strength.
It is also possible to optimize the setup with respect to the cavity mode width and explore the dependence $n_{0}\sim \xi^{-6/5}$.

\subsection{Thermal effects}
In any experiment, the thermal pressure of the Bose gas $P_{\rm th} \propto T^4(gn)^{-3/2}$ [see Eq.\ (\ref{thermalPressure})] will contribute by shifting the equilibrium point to a lower density.
Therefore, an optimization of the experiment should focus on enabling a denser droplet.
At nanokelvin temperatures, a system of $N=100$ atoms and an envelope waist $\xi = 5 \, \mu$m can sustain the droplet close to the Dicke phase transition $\mathcal{I} = 0.99 \, \mathcal{I}_{\rm cr}$.
To get a stable droplet with the same parameters as in Fig.\ \ref{energyPerParticle}, the condensate must be at $T < 0.1$ nK.
Condensates have been prepared at such low temperatures \cite{leanhardt2003cooling} even at tens of picokelvin with matter wave lensing techniques \cite{deppner2021collective,gaaloul2022space}.
Increasing the interaction strength $g$ proves effective in suppressing detrimental thermal effects. 
While finite temperatures may even be beneficial to promote quantum effects in other types of droplets \cite{sanchez2023heating}, they hinder the formation of cavity quantum droplets of type (D3).
Due to $P_{\rm th}\propto T^4$, the temperature window is narrow for a droplet, where $P_{\rm th}$ is significant.

\section{Conclusions}
We have shown that a mean-field stable BEC can be turned into a stable quantum droplet when coupled to a single-mode cavity with transverse pumping.
The zero-point energy of the cavity quantum fluctuations then provides an attractive quantum correction to the mean-field BEC contact interaction energy if the cavity is red detuned.
The dependence of the attractive term on the BEC volume is crucially influenced by the shape of the cavity mode function, which determines the cavity-induced atom-atom interaction.
The Fourier transform of the mode function envelope determines the roton eigenfrequency and thus the finite-size behavior of the roton energy contribution.
An alternative view is that the cavity quantum fluctuations, once strongly coupled to the BEC, exert a significant negative pressure on the atomic cloud.
It turns out that the spatial profiles of the light fields inducing the finite-range interaction are essential for droplet formation.
The equilibrium density of cavity-induced quantum droplets can be controlled by varying the strength of the transverse pump or the cavity detuning.
Self-boundedness is a key experimental hallmark of these quantum objects. 
If a condensate is prepared, coupled to the cavity, and then released by removing the external trapping along at least one axis, absorption imaging could reveal whether the cloud maintains its form or expands.
\\
\indent
Although we have focused on the simplest form of a cavity-induced long-range interaction, the theory can account for any envelope function $f(\bm{r},\bm{r}')$ and system dimension.
Our theory, e.g., can directly be applied \cite{companionPaper} to multimode cavities with an envelope $f(x, x')=\exp[-(x-x')^2/\xi^2]$ \cite{mivehvar2021cavity}. Those have been studied in a lattice with quantum Monte Carlo techniques \cite{karpov2022light} and in terms of numerical solutions for ring cavities \cite{masalaeva2023tuning}.
\\
\indent

\section*{Acknowledgments}
We are grateful to Andreas Hemmerich and Hans Keßler for helpful discussions on the experimental parameters. This work was supported by the Deutsche Forschungsgemeinschaft (DFG, German Research Foundation) via the Collaborative Research Center SFB/TR185 (Project No.\ 277625399) (A.P.) and via the Research Grant No.\ 274978739 (M.R.\ and M.T.). We also acknowledge the support from the DFG Cluster of Excellence CUI: ``Advanced Imaging of Matter'' -- EXC 2056 (Project No. 390715994).

\appendix
%\newpage

% \AddAppendix
\section{Approximations leading to Eq.\ (\ref{KroneckerDelta})\label{App:A}}
When the homogeneous mean-field approximation $\hat{\psi}(\bm{r})\to\langle \hat{\psi}(\bm{r})\rangle = \sqrt{n}$ is used in the second line of Eq.\ (\ref{effectiveInteractingBEC}), one finds
\begin{align}
\frac{In^2}{32} \bigg[ \int_{V} d^3\bm{r}\;\! \big( e^{ikx} + e^{-ikx} \big) \big( e^{iky} + e^{-iky} \big)\;\!e^{- (y^2+z^2)/\xi^2}\bigg]^2 \, .
\label{spatialInteractionAverage}
\end{align}
Since the atoms occupy a cube of volume $V=L^3$, it decomposes into the product of integrals given in Eq.\ (\ref{KroneckerDelta}).
For discrete momenta $k = 2\pi m/L$ for $m \in \mathbb{Z}$ and for the constraint $L < \xi$, the Gaussian $e^{-u^2/\xi^2}$ varies slowly compared to the fast oscillating exponential $e^{iku}$ for $k\neq 0$.
Hence, we can approximate the Gaussian envelope by its spatial average
\begin{align}
\int_{-L/2}^{L/2} du\;\! e^{iku} e^{-u^2/\xi^2} 
&\approx \int_{-L/2}^{L/2} du\;\! e^{iku} \int_{-L/2}^{L/2} \frac{du}{L} e^{-u^2/\xi^2}\nonumber\displaybreak[0] \\
&= \delta_{k0} \sqrt{\pi} \xi \erf\bigg( \frac{L}{2\xi} \bigg) \, .
\label{spatialAverageApprox}
\end{align}
In this approximation, the spatial integral in Eq.\ (\ref{spatialInteractionAverage}) vanishes since $k\neq 0$.

% \AddAppendix
\section{Bogoliubov theory\label{App:B}}
To include quantum fluctuations, we expand the Hamiltonian in Eq.\ (\ref{effectiveInteractingBEC}) up to the second order in $\hat{\phi}(\bm{r}) = \hat{\psi}(\bm{r}) - \sqrt{n}$. The linear order vanishes upon applying Eq.\ (\ref{spatialAverageApprox}). The quadratic order reads
\vspace{-4mm}
\begin{align}
\hat{H}_{\rm qf} = &\int_V d^3\bm{r}\;\! \left[ \hat{\phi}^{\dag} \bigg( -\frac{\mbox{$\bm{\nabla}^2$}}{2M} + gn \bigg) \hat{\phi} + \frac{gn}{2} \Big( \hat{\phi}\hat{\phi} + \hat{\phi}^{\dag}\hat{\phi}^{\dag} \Big) \right] \nonumber \\
&+ \frac{\mathcal{I}n}{2} \bigg[ \int_V d^3\bm{r} \Big( \hat{\phi} + \hat{\phi}^{\dag} \Big) \cos(kx)\cos(ky)\;\! e^{-\frac{y^2+z^2}{\xi^2}} \bigg]^2 \, ,
\end{align}
where we suppressed the spatial dependence of the fluctuation operators $\hat{\phi}$ and $\hat{\phi}^{\dag}$ for brevity.
The plane wave expansion of the fluctuation operator $\hat{\phi}(\bm{r}) = \sum_{\bm{p}\neq \bm{0}} \hat{\phi}_{\bm{p}} e^{i\bm{p}\bm{r}}/\sqrt{V}$ leads to
\begin{widetext}
\begin{align}
\hat{H}_{\rm qf} ={}&{} \sum_{\bm{p},\bm{p}'}{}^{'} \hat{\phi}_{\bm{p}}^{\dag} \bigg( \frac{\bm{p}'^2}{2M} + gn \bigg) \hat{\phi}_{\bm{p}'} \int_V \frac{d^3\bm{r}}{V} e^{-i(\bm{p}-\bm{p}')\bm{r}} + \bigg( \frac{gn}{2} \hat{\phi}_{\bm{p}} \hat{\phi}_{\bm{p}'} \int_V \frac{d^3\bm{r}}{V} e^{i(\bm{p}+\bm{p}')\bm{r}} + h.c. \bigg) \nonumber \\
&+ \frac{\mathcal{I}N}{2} \bigg\{ \sum_{\bm{p}}{}^{'} \bigg[ \hat{\phi}_{\bm{p}} \int_V \frac{d^3 \bm{r}}{V} \cos(kx) \cos(ky) e^{-i\bm{p}\bm{r}-\frac{y^2+z^2}{\xi^2}} + h.c. \bigg] \bigg\}^2 \, ,
\end{align}
\end{widetext}
where the primed sum indicates the omission of the zero momentum terms.
The integrals are performed using the Kronecker integral representation $\int_{-L/2}^{L/2} du\;\! e^{ipu} = L \delta_{p0}$ and Eq.\ (\ref{spatialAverageApprox}), yielding
\begin{align}
\hspace*{-2mm}\hat{H}_{\rm qf} &= \hspace*{-1mm}\sum_{\bm{p}}{}^{'}\left[ \bigg( \frac{\bm{p}^2}{2M} + gn \bigg) \hat{\phi}_{\bm{p}}^{\dag} \hat{\phi}_{\bm{p}} + \frac{gn}{2} \Big(\hat{\phi}_{-\bm{p}} \hat{\phi}_{\bm{p}} + \hat{\phi}_{\bm{p}}^{\dag} \hat{\phi}_{-\bm{p}}^{\dag} \Big) \right]\nonumber \\
&\hspace*{-5mm}+ \frac{\mathcal{I}N}{32V^2} \bigg\{ \bigg[ \sqrt{\pi} \xi \erf\bigg( \frac{L}{2\xi} \bigg) \bigg]^2 \sum_{\bm{k}\in\mathcal{K}_C} \Big(\hat{\phi}_{\bm{k}} + \hat{\phi}_{\bm{k}}^{\dag} \Big)\bigg\}^2 \, ,
\end{align}
with \mbox{$\mathcal{K}_{\rm C} = \lbrace ({\!}\pm{\;\!\!}k\;\:\pm{\;\!\!}k\;\: 0 )^T \rbrace$}.
Its eigenmodes are obtained by a standard Bogoliubov transformation to the quasiparticles of the modes unaffected by the cavity-induced interaction $\bm{p}\notin\mathcal{K}_C$. An additional transformation is required for those $\bm{k}\in\mathcal{K}_C$ that are coupled to the cavity.
The former follow the standard Bogoliubov dispersion 
\begin{align}
\omega_{\bm{p}} = \sqrt{\frac{\bm{p}^2}{2M} \bigg( \frac{\bm{p}^2}{2M} + 2gn \bigg)} \, .
\label{end:boseGasDispersion}
\end{align}
To determine the eigenmodes for $\bm{k}\in\mathcal{K}_C$ it is convenient to express the Hamiltonian in terms of the quasi-position $\hat{x}_{\bm{p}} = \sqrt{M}(\hat{\phi}_{\bm{p}} + \hat{\phi}_{-\bm{p}}^{\dag})/\sqrt{\bm{p}^2}$ and quasi-momentum $\hat{y}_{\bm{p}} = -i \sqrt{\bm{p}^2}(\hat{\phi}_{-\bm{p}} - \hat{\phi}_{\bm{p}}^{\dag})/\sqrt{4M}$ operators of Ref.\ \cite{tommasini2003bogoliubov}, yielding 
\begin{align}
\hat{H}_{\rm qf} ={}&{}  \frac{1}{2} \sum_{\bm{p}\notin\mathcal{K}_C}{}^{'} \bigg( \hat{y}_{\bm{p}}^{\dag} \hat{y}_{\bm{p}} + \omega_{\bm{p}}^2 \hat{x}_{\bm{p}}^{\dag} \hat{x}_{\bm{p}} - \frac{\bm{p}^2}{2M} - gn \bigg) \nonumber \\
&+{} \frac{1}{2} \bigg[ \hat{\vec{y}\;\!}^{\dag} \mathbb{I}_{4\times 4}\:\! \hat{\vec{y}} + \hat{\vec{x}\;\!}^{\dag} \underline{h}\;\! \hat{\vec{x}} - 4 \bigg( \frac{\bm{k}^2}{2M} + gn \bigg) \bigg] \, .
\label{qfHamiltonian}
\end{align}
Here we have defined \mbox{$\hat{\vec{x}} = (\hat{x}_{\bm{k}_1}\;\:\hat{x}_{\bm{k}_2}\;\:\hat{x}_{\bm{k}_3}\;\:\hat{x}_{\bm{k}_4} )^T$} and $\hat{\vec{y}}$ analogously, where $\mathcal{K}_C = \lbrace \bm{k}_1,\bm{k}_2,\bm{k}_3,\bm{k}_4 \rbrace$.
The cavity-mediated interaction in Eq.\ (\ref{cavityInteractionPotential}) only couples quasi-position operators so that the eigenmodes can be read off from the eigenvalues of the $4\times 4$ matrix coupling the four modes in $\mathcal{K}_C$
\begin{align}
\underline{h} = \omega_{\bm{k}}^2\;\! \mathbb{I}_{4\times 4} + \frac{\mathcal{I} N \bm{k}^2}{16 M} \bigg[ \frac{\sqrt{\pi}\xi}{L} \erf\bigg( \frac{L}{2\xi} \bigg) \bigg]^4 \mathbb{J}_{4\times 4} \, ,
\label{couplingMatrix}
\end{align}
where $\mathbb{J}_{4\times 4}$ is the matrix of all elements $1$.
The eigenvalues are $\{ \omega_{\bm{k}},\omega_{\bm{k}},\omega_{\bm{k}},\Omega \}$ with the first three obeying the standard Bogoliubov dispersion $\omega_{\bm{k}}$ 
from Eq.\ (\ref{end:boseGasDispersion})
and the last being affected by the cavity according to Eq.\ (\ref{rotonMode}).
Using these eigenmodes, we find the quantum fluctuations zero-point energy
\begin{align}
E_{\rm qf} = \frac{1}{2} \sum_{\bm{p}}{}^{'} \bigg( \omega_{\bm{p}} - \frac{\bm{p}^2}{2M} - gn \bigg) + \frac{1}{2} \big( \Omega - \omega_{\bm{k}} \big) \, .
\label{quantumCorrection}
\end{align}
The first term is the LHY correction of a Bose gas \cite{lee1957eigenvalues} that is to be evaluated in the continuum limit with the proper renormalization and equals $8VM^{3/2}(gn)^{5/2}/(15\pi^2)$.
In a weakly interacting dilute Bose gas, as studied here, the LHY correction is negligible compared to the mean-field energy in Eq.\ (\ref{meanFieldPotential}).
The second term encapsulates the deviation of the roton dispersion from the Bogoliubov one due to the presence of the cavity-mediated interaction.
It is exactly the cavity-induced quantum fluctuation correction $E_{\rm ac}$ given in Eq.\ (\ref{cavityCorrection}).

% \AddAppendix

% \section{Density and interaction strength\label{App:C}}
% Figure \ref{DensityInteraction} displays the droplet density as a function of the atom number and the system size.
% The magenta lines correspond to the zero-pressure contours of long-range interaction strength values, $\mathcal{I}$, ranging from $0.1\,\mathcal{I}_{\rm cr}$ to $0.95\,\mathcal{I}_{\rm cr}$, from the top left to the lower right corner as a geometric sequence.
% The magenta border line is the same as the magenta line in Fig.\ \ref{pressureDiagram}.
% Recall that $\mathcal{I}_{\rm cr} \sim N^{-1}$ depends on the number of atoms.
% The solid cyan lines correspond to zero-pressure contours with {\it constant} values of $|\mathcal{I}| \approx 9\, \text{ Hz\ldash{}}\, 515$ Hz, which are chosen to match the dashed curves for $N = 100$.
    
% \AddAppendix

\section{Derivation of the thermal pressure\label{App:D}}
At  finite temperature,   the effective energy $E_0(N,V,T) = E_{\rm mf} + E_{\rm ac} + E_{\rm th}$ acquires an additional thermal contribution \cite{pitaevskii,sanchez2023heating}
\begin{align}
\hspace*{-2mm}E_{\rm th} = \frac{1}{\beta} \sum_{\bm{p}}{}^{'} \ln \big( 1 - e^{-\beta\omega_{\bm{p}}} \big) + \frac{1}{\beta} \ln \bigg( \frac{1-e^{-\beta\Omega}}{1-e^{-\beta\omega_{\bm{k}}}} \bigg) .
\label{thermalContribution}
\end{align}
The sum over the momenta is evaluated as an integral in the continuum limit.
The ultra-low temperature regime implies large $\beta$, such that the Bogoliubov dispersion can be approximated as phononic, i.e., $\ln (1 - e^{-\beta\omega_{\bm{p}}}) \approx \ln( 1 - e^{-\beta \sqrt{gn\bm{p}^2/M}})$ \cite{pitaevskii}.
This gives
\begin{align}
E_{\rm th} = -\frac{\pi^2 V M^{3/2}}{90\beta^4(gn)^{3/2}} + \frac{1}{\beta} \ln \bigg( \frac{1-e^{-\beta\Omega}}{1-e^{-\beta\omega_{\bm{k}}}} \bigg) \, .
\label{lowTthermalTerm}
\end{align}
The resulting thermal pressure is then
\begin{align}
P_{\rm th} = \frac{\pi^2 M^{3/2}}{36\beta^4(gn)^{3/2}} \, ,
\label{thermalPressure}
\end{align}
where the remaining roton contribution is negligible.


\begin{thebibliography}{100}
%
\bibitem{kapitza1938viscosity} 
P. Kapitza, 
\textit{Viscosity of liquid helium below the $\lambda$-point}, \href{https://doi.org/10.1038/141074a0}{Nature {\bf 141}, 74 (1938)}.
%
\bibitem{landau1941theory} 
L. Landau, 
\textit{Theory of the superfluidity of helium II}, 
\href{https://doi.org/10.1103/PhysRev.60.356}{Phys. Rev. {\bf 60}, 356 (1941)}.
%
\bibitem{volovik2003universe} 
G. E. Volovik, 
{\it The universe in a helium droplet} 
(Oxford University Press, Oxford, 2003).
%
\bibitem{stringari1987systematics}
S. Stringari and J. Treiner, \textit{Systematics of liquid helium clusters}, \href{https://doi.org/10.1063/1.452818}{J. Chem. Phys. {\bf 87}, 5021 (1987)}.
%
\bibitem{dupont1990inhomogeneous}
J. Dupont-Roc, M. Himbert, N. Pavloff, and J. Treiner, \textit{Inhomogeneous liquid ${}^4$He: A density functional approach with a finite-range interaction}, \href{https://doi.org/10.1007/BF00683150}{J. Low Temp. Phys. {\bf 81}, 31 (1990)}.
%
\bibitem{casas1995density} 
M.\ Casas, F.\ Dalfovo, A.\ Lastri, Ll.\ Serra, and S.\ Strin\-gari, \textit{Density functional calculations for ${}^4$He droplets}, 
\href{https://doi.org/10.1007/BF01439984}{Z. Phys. D {\bf 35}, 67 (1995)}.
%
\bibitem{dalfovo1995structural} 
F. Dalfovo, A. Lastri, L. Pricaupenko, S. Stringari, and J. Treiner, 
\textit{Structural and dynamical properties of superfluid helium: A density-functional approach}, 
\href{https://doi.org/10.1103/PhysRevB.52.1193}{Phys. Rev. B {\bf 52}, 1193 (1995)}.
%
\bibitem{petrov2015quantum} 
D. S. Petrov, 
\textit{Quantum mechanical stabilization of a collapsing Bose-Bose mixture}, 
\href{https://doi.org/10.1103/PhysRevLett.115.155302}{Phys. Rev. Lett. {\bf 115}, 155302 (2015)}.
%
\bibitem{lee1957eigenvalues} 
T. D. Lee, K. Huang, and C. N. Yang, \textit{Eigenvalues and eigenfunctions of a Bose system of hard spheres and its low-temperature properties}, 
\href{https://doi.org/10.1103/PhysRev.106.1135}{Phys. Rev. {\bf 106}, 1135 (1957)}.
%
\bibitem{cabrera2018quantum} 
C. R. Cabrera, L. Tanzi, J. Sanz, B. Naylor, P. Thomas, P. Cheiney, and L. Tarruell, \textit{Quantum liquid droplets in a mixture of Bose-Einstein condensates}, 
\href{https://www.science.org/doi/10.1126/science.aao5686}{Science {\bf 359}, 301 (2018)}.
%
\bibitem{semeghini2018self} 
G. Semeghini, G. Ferioli, L. Masi, C. Mazzinghi, L. Wolswijk, F. Minardi, M. Modugno, G. Modugno, M. Inguscio, and M. Fattori, 
\textit{Self-bound quantum droplets of atomic mixtures in free space}, 
\href{https://doi.org/10.1103/PhysRevLett.120.235301}{Phys. Rev. Lett. {\bf 120}, 235301 (2018)}.
%
\bibitem{skov2021observation} 
T. G. Skov, M. G. Skou, N. B. J{\o}rgensen, and J. J. Arlt, 
\textit{Observation of a Lee-Huang-Yang fluid}, 
\href{https://doi.org/10.1103/PhysRevLett.126.230404}{Phys. Rev. Lett. {\bf 126}, 230404 (2021)}.
%
\bibitem{ferrier2016observation} 
I. Ferrier-Barbut, H. Kadau, M. Schmitt, M. Wenzel, and T. Pfau, 
\textit{Observation of quantum droplets in a strongly dipolar Bose gas}, 
\href{https://doi.org/10.1103/PhysRevLett.116.215301}{Phys. Rev. Lett. {\bf 116}, 215301 (2016)}.
%
\bibitem{pfau-nature}
M. Schmitt, M. Wenzel, F. Böttcher, I. Ferrier-Barbut, and T. Pfau,
\textit{Self-bound droplets of a dilute magnetic quantum liquid},
\href{https://doi.org/10.1038/nature20126}{Nature \textbf{539}, 259 (2016)}.
%
\bibitem{chomaz2022dipolar} 
L. Chomaz, I. Ferrier-Barbut, F. Ferlaino, B. Laburthe-Tolra, B. L. Lev, and T. Pfau, \textit{Dipolar physics: A review of experiments with magnetic quantum gases}, 
\href{https://iopscience.iop.org/article/10.1088/1361-6633/aca814}{Rep. Prog. Phys. {\bf 86}, 026401 (2023)}.
%
\bibitem{schutzhold2006mean}
R. Sch{\"u}tzhold, M. Uhlmann, Y. Xu, and U. R. Fischer, 
\textit{Mean-field expansion in Bose-Einstein condensates with finite-range interactions},
\href{https://doi.org/10.1142/S0217979206035631}{Int. J. Mod. Phys. B {\bf 20}, 3555 (2006)}.
%
\bibitem{lima2011quantum} 
A. R. P. Lima and A. Pelster, 
\textit{Quantum fluctuations in dipolar Bose gases}, 
\href{https://doi.org/10.1103/PhysRevA.84.041604}{Phys. Rev. A {\bf 84}, 041604(R) (2011)}.
%
\bibitem{lima2012beyond} 
A. R. P. Lima and A. Pelster, 
\textit{Beyond mean-field low-lying excitations of dipolar Bose gases}, 
\href{https://doi.org/10.1103/PhysRevA.86.063609}{Phys. Rev. A {\bf 86}, 063609 (2012)}.
%
\bibitem{wachtler2016quantum} 
F. W{\"a}chtler and L. Santos, 
\textit{Quantum filaments in dipolar Bose-Einstein condensates}, 
\href{https://doi.org/10.1103/PhysRevA.93.061603}{Phys. Rev. A {\bf 93}, 061603(R) (2016)}.
%
\bibitem{bisset2016ground} 
R. N. Bisset, R. M. Wilson, D. Baillie, and P. B. Blakie, 
\textit{Ground-state phase diagram of a dipolar condensate with quantum fluctuations}, 
\href{https://doi.org/10.1103/PhysRevA.94.033619}{Phys. Rev. A {\bf 94}, 033619 (2016)}.
%
\bibitem{roton-prediction1}
L. Santos, G. V. Shlyapnikov, and M. Lewenstein, 
\textit{Roton–maxon spectrum and stability of trapped dipolar Bose-Einstein condensates}, \href{https://doi.org/10.1103/PhysRevLett.90.250403}{Phys. Rev. Lett. \textbf{90},
250403 (2003)}.
%
\bibitem{roton-prediction2}
D. H. J. O’Dell, S. Giovanazzi, and G. Kurizki, 
\textit{Rotons in gaseous Bose-Einstein condensates irradiated by a laser},
\href{https://journals.aps.org/prl/abstract/10.1103/PhysRevLett.90.110402}{Phys. Rev. Lett. \textbf{90}, 110402 (2003)}.
%
\bibitem{chomaz2018observation} 
L. Chomaz, R. M. W. van Bijnen, D. Petter, G. Faraoni, S. Baier, J. H. Becher, M. J. Mark, F. Waechtler, L. Santos, and F. Ferlaino, 
\textit{Observation of roton mode population in a dipolar quantum gas}, 
\href{https://doi.org/10.1038/s41567-018-0054-7}{Nat. Phys. {\bf 14}, 442 (2018)}.
%
\bibitem{petter2019probing}
D. Petter, G. Natale, R. M. W. van Bijnen, A. Patscheider, M. J. Mark, L. Chomaz, and F. Ferlaino, 
\textit{Probing the roton excitation spectrum of a stable dipolar Bose gas}, 
\href{https://doi.org/10.1103/PhysRevLett.122.183401}{Phys. Rev. Lett. {\bf 122}, 183401 (2019)}.
%
\bibitem{Petrov2016} D. S. Petrov and G. E. Astrakharchik, \textit{Ultradilute low-dimensional liquids}, 
\href{https://doi.org/10.1103/PhysRevLett.117.100401}{Phys. Rev. Lett. {\bf 117}, 100401 (2016)}.
%
\bibitem{Ilg}
T. Ilg, J. Kumlin, L. Santos, D. S. Petrov, and H. P. B{\"u}chler,
\textit{Dimensional crossover for the beyond-mean-field correction in Bose gases},
\href{https://doi.org/10.1103/PhysRevA.98.051604}{Phys. Rev. A \textbf{98}, 051604(R) (2018)}.
%
\bibitem{jia2022expansion} 
F. Jia, Z. Huang, L. Qiu, R. Zhou, Y. Yan, and D. Wang, 
\textit{Expansion dynamics of a shell-shaped Bose-Einstein condensate}, 
\href{https://doi.org/10.1103/PhysRevLett.129.243402}{Phys. Rev. Lett. {\bf 129}, 243402 (2022)}.
%
\bibitem{dicke1954coherence} R. H. Dicke, \textit{Coherence in spontaneous radiation processes},\href{https://doi.org/10.1103/PhysRev.93.99}{Phys. Rev. {\bf 93}, 99 (1954)}.
%
\bibitem{nagy2010dicke} 
D. Nagy, G. K{\'o}nya, G. Szirmai, and P. Domokos, 
\textit{Dicke-model phase transition in the quantum motion of a Bose-Einstein condensate in an optical cavity}, 
\href{https://doi.org/10.1103/PhysRevLett.104.130401}{Phys. Rev. Lett. {\bf 104}, 130401 (2010)}.
%
\bibitem{baumann2010dicke} 
K. Baumann, C. Guerlin, F. Brennecke, and T. Esslinger, 
\textit{Dicke quantum phase transition with a superfluid gas in an optical cavity}, 
\href{https://doi.org/10.1038/nature09009}{Nature {\bf 464}, 1301 (2010)}.
%
\bibitem{mivehvar2021cavity} 
F. Mivehvar, F. Piazza, T. Donner, and H. Ritsch, 
\textit{Cavity QED with quantum gases: new paradigms in many-body physics}, 
\href{https://www.tandfonline.com/doi/full/10.1080/00018732.2021.1969727}{Adv. Phys. {\bf 70}, 1 (2021)}.
%
\bibitem{emary2003chaos} C. Emary, and T. Brandes, \textit{Chaos and the quantum phase transition in the Dicke model}, \href{https://doi.org/10.1103/PhysRevE.67.066203}{Phys. Rev. E {\bf 67}, 066203 (2003)}.
%
\bibitem{mottl2012roton} 
R. Mottl, F. Brennecke, K. Baumann, R. Landig, T. Donner, and T. Esslinger, 
\textit{Roton-type mode softening in a quantum gas with cavity-mediated long-range interactions}, 
\href{https://www.science.org/doi/10.1126/science.1220314}{Science {\bf 336}, 1570 (2012)}.
%
\bibitem{karpov2019crystalline}
P. Karpov and F. Piazza,
\textit{Crystalline droplets with emergent color charge in many-body systems with sign-changing interactions},
\href{https://doi.org/10.1103/PhysRevA.100.061401}{Phys. Rev. A {\bf 100}, 061401 (2019)}.
%
\bibitem{karpov2022light} 
P. Karpov and F. Piazza, 
\textit{Light-induced quantum droplet phases of lattice bosons in multimode cavities}, \href{https://doi.org/10.1103/PhysRevLett.128.103201}{Phys. Rev. Lett. {\bf 128}, 103201 (2022)}.
%
\bibitem{companionPaper}
L. Mixa, M. Radonjić, A. Pelster, and M. Thorwart, 
\textit{Engineering quantum droplet formation by cavity-induced long-range interactions}, \href{https://doi.org/10.1103/PhysRevResearch.7.023204}{Phys. Rev. Res. {\bf 7}, 023204 (2025)}.
%
\bibitem{maschler2008ultracold} 
C. Maschler, I. B. Mekhov, and H. Ritsch, \textit{Ultracold atoms in optical lattices generated by quantized light fields}, 
\href{https://link.springer.com/article/10.1140/epjd/e2008-00016-4}{Eur. Phys. J. D {\bf 46}, 545 (2008)}.
%
\bibitem{jager2022lindblad} 
S. B. J\"ager, T. Schmit, G. Morigi, M. J. Holland, and R. Betzholz, 
\textit{Lindblad master equations for quantum systems coupled to dissipative bosonic modes}, 
\href{https://doi.org/10.1103/PhysRevLett.129.063601}{Phys. Rev. Lett. {\bf 129}, 063601 (2022)}.
%
\bibitem{nagy2011critical} D. Nagy, G. Szirmai, and P. Domokos, \textit{Critical exponent of a quantum-noise-driven phase transition: The open-system Dicke model}, \href{https://doi.org/10.1103/PhysRevA.84.043637}{Phys. Rev. A {\bf 84}, 043637 (2011)}.
%
\bibitem{gajda2021stability}
P. Zin, M. Pylak, and M. Gajda,
\textit{Revisiting a stability problem of two-component quantum droplets},
\href{https://doi.org/10.1103/PhysRevA.103.013312}{Phys. Rev. A {\bf 103}, 013312 (2021)}.
%
\bibitem{leanhardt2003cooling}
A. E. Leanhardt, T. A. Pasquini, M. Saba, A. Schirotzek, Y. Shin, D. Kielpinski, D. E. Pritchard, and W. Ketterle,
\textit{Cooling Bose-Einstein condensates below 500 picokelvin}, 
\href{https://doi.org/10.1126/science.1088827}{Science {\bf 301}, 1513 (2003)}.
%
\bibitem{deppner2021collective}
C. Deppner et al., 
\textit{Collective-mode enhanced matter-wave optics},
\href{https://doi.org/10.1103/PhysRevLett.127.100401}{Phys. Rev. Lett. {\bf 127}, 100401 (2021)}.
%
\bibitem{gaaloul2022space}
N. Gaaloul et al.,
\textit{A space-based quantum gas laboratory at picokelvin energy scales}, 
\href{https://doi.org/10.1038/s41467-022-35274-6}{Nat. Commun. {\bf 13}, 7889 (2022)}.
%
\bibitem{sanchez2023heating}
J. Sánchez-Baena, C. Politi, F. Maucher, F. Ferlaino, and T. Pohl, 
\textit{Heating a dipolar quantum fluid into a solid}, 
\href{https://doi.org/10.1038/s41467-023-37207-3}{Nat. Commun. {\bf 14}, 1868 (2023)}.
%
\bibitem{masalaeva2023tuning} 
N. Masalaeva, H. Ritsch, and F. Mivehvar, \textit{Tuning photon-mediated interactions in a multimode cavity: From supersolid to insulating droplets hosting phononic excitations}, 
\href{https://doi.org/10.1103/PhysRevLett.131.173401}{Phys. Rev. Lett. {\bf 131}, 173401 (2023)}.
%
\bibitem{tommasini2003bogoliubov}
P. Tommasini, E. J. V. de Passos, A. F. R. de Toledo Piza, M. S. Hussein, and E. Timmermans,
\textit{Bogoliubov theory for mutually coherent condensates},
\href{https://doi.org/10.1103/PhysRevA.67.023606}{Phys. Rev. A \textbf{67}, 023606 (2003)}.
%
\bibitem{pitaevskii}
L. Pitaevskii and S. Stringari,
\textit{Bose-Einstein condensation and superfluidity}
(Oxford University Press, Oxford, 2016).
%
\end{thebibliography}
\end{document}